RESEARCH

# PhyLiNO: a forward-folding likelihood-fit framework for neutrino oscillation physics

Denise Hellwig[1,3] · Stefan Schoppmann[1,2] · Philipp Soldin[1] · Achim Stahl[1] · Christopher Wiebusch[1]

**Abstract**
We present a framework for the analysis of data from neutrino oscillation experiments. The framework performs a profile likelihood fit and employs a forward-folding technique to optimize its model with respect to the oscillation parameters. It is capable of simultaneously handling multiple datasets from the same or different experiments and their correlations. The code of the framework is optimized for performance and allows for convergence times of a few seconds handling hundreds of fit parameters, thanks to multi-threading and usage of GPUs. The framework was developed in the context of the Double Chooz experiment, where it was successfully used to fit three- and four-flavor models to the data, as well as in the measurement of the energy spectrum of reactor neutrinos. We demonstrate its applicability to other experiments by applying it to a study of the oscillation analysis of a medium baseline reactor experiment similar to JUNO.

**Keywords** Neutrino oscillations · Likelihood fit · Forward folding

## Introduction

In neutrino physics, a common task is extracting oscillation parameters such as mixing angles $\theta_{jk}$ and squared mass differences $\Delta m^2_{jk}$, with $j, k \in \{1, 2, 3\}$, $j \neq k$, by fitting the theoretical model to a measured dataset in a Neutrino detector [1–3]. The experimental data will have detector effects folded through, smearing the true underlying distribution. Inverting this by correcting the acquired data for detector effects in order to compare it to the prediction, the so-called unfolding, is often challenging. An alternative is forward-folding which is based on predicting the experimental data using a model of the detector response and fitting this to the acquired data. This approach usually requires significant computing resources, as the model expectation has to be recalculated frequently during the fitting procedure.

In this article, we present a software framework for forward-folding fits for use in neutrino oscillation physics and potentially beyond. It exploits a binned log-likelihood approach that allows combining datasets from multiple detectors and varying operational conditions. It accounts for correlated systematic uncertainties of the expected rates and spectra of signal and background contributions, as well as the energy scale and other detector uncertainties. External knowledge on systematic effects can be included by virtue of prior functions.

For practical use, the algorithms are optimized for computational efficiency. The framework performs a full oscillation fit with $\mathcal{O}(300)$ parameters within roughly a minute on a typical notebook. Applying techniques of parallelization and using Graphics Processing Units (GPUs), improves the convergence time to a few seconds. With this, large ensemble studies of simulated pseudo-experiments, e.g., for validating the statistical coverage of results, become feasible.

Before we discuss details of our framework in section Details of the framework, we introduce the experimental context in section Experimental Context. We discuss extensions and possible applications of our framework in section Extensions of the framework, and conclude with a

✉ Stefan Schoppmann
schoppmann@uni-mainz.de

Philipp Soldin
soldin@physik.rwth-aachen.de

1 III. Physikalisches Institut, RWTH Aachen University, 52056 Aachen, Germany

2 Detektorlabor, Exzellenzcluster PRISMA+, Johannes Gutenberg-Universität Mainz, 55128 Mainz, Germany

3 Present Address: d-fine GmbH, 60313 Frankfurt am Main, Germany

summary in section Conclusion. We provide the source code of our framework through a GitHub repository [4].

## Experimental context

The software framework presented in this article has been developed in the context of the reactor anti-neutrino disappearance experiment Double Chooz [5, 6]. Double Chooz is a short-baseline reactor anti-neutrino disappearance experiment. It consists of two almost-identical liquid scintillator detectors measuring the neutrino flux from two nuclear reactors. The near detector (ND) and far detector (FD) have an approximate distance of 400 m and 1050 m to the reactor cores, respectively. The FD started operation in 2011 (FD-I dataset) and since 2015 both detectors were operational (FD-II and ND datasets). Data taking concluded in December 2017. Double Chooz's primary goal is the measurement of the survival probability $P_{\bar{\nu}_e \rightarrow \bar{\nu}_e}$ of electron anti-neutrinos and thereby the measurement of the lepton mixing angle $\theta_{13}$ [2, 7]. The survival probability for a short-baseline reactor experiment is given in good approximation as:

$$
\begin{aligned}
P_{\bar{\nu}_e \rightarrow \bar{\nu}_e}(L, E) \approx & 1 - \sin^2(2\theta_{13}) \sin^2\left(\frac{\Delta m^2_{ee} L}{4E}\right) \\
& - \cos^4(\theta_{13}) \sin^2(2\theta_{12}) \sin^2\left(\frac{\Delta m^2_{21} L}{4E}\right),
\end{aligned}
\tag{1}
$$

where $L$ is the baseline from the reactor to the detector, $E$ the neutrino energy, $\theta_{12}$ and $\theta_{13}$ are two of the leptonic mixing angles, $\Delta m^2_{ij}$ is a squared mass splitting, and

$$
\Delta m^2_{ee} \equiv \cos^2(\theta_{12}) |\Delta m^2_{31}| + \sin^2(\theta_{12}) |\Delta m^2_{32}|
\tag{2}
$$

is the effective squared mass splitting at reactors [8]. The main objective of our software framework has been the determination of the oscillation parameters, primarily $\theta_{13}$, in Eq. 1 considering all experimental and theoretical uncertainties.

While the primary purpose of this software framework in Double Chooz has been the determination of the mixing angle $\theta_{13}$, see [9] and the upcoming result [10], it has further been used by the Double Chooz collaboration to investigate the shape anomaly of the reactor spectrum [9] and in the search for sterile neutrinos [11, 12].

### Neutrino Signal

Reactor neutrinos can be efficiently detected by the reaction of the inverse $\beta$-decay (IBD) [6]. In this reaction, an electron anti-neutrino $\bar{\nu}_e$ interacts with the scintillator and converts into a positron and a neutron. The positron

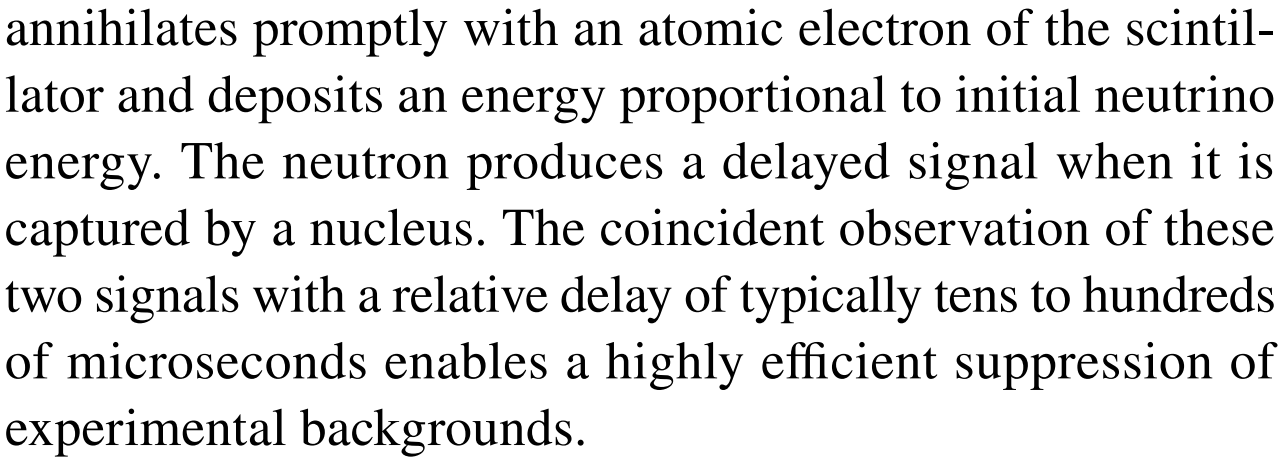
annihilates promptly with an atomic electron of the scintillator and deposits an energy proportional to initial neutrino energy. The neutron produces a delayed signal when it is captured by a nucleus. The coincident observation of these two signals with a relative delay of typically tens to hundreds of microseconds enables a highly efficient suppression of experimental backgrounds.

The expected rate and spectrum of anti-neutrinos is estimated from reactor and detector-specific inputs such as detection efficiency, fiducial mass, baseline, thermal power of each reactor, energy release per fission of the fuel isotopes, and the time-dependent fuel inventory of each reactor core [13]. These estimations enter the fit only indirectly as starting values and for the computation of covariances of bin-to-bin normalizations. Due to the two-detector setup, where the measured experimental rates are compared, the overall normalizations as well as the shape of the energy spectrum are determined from the measured data. Here, the ND provides a reference measurement of the almost un-oscillated neutrino flux. When comparing to the FD measurement, the difference to the un-oscillated prediction is due to neutrino oscillations and is described by Eq. 1.

### Background processes

There are few background processes which produce a signature similar to an IBD. These are either correlated processes induced by atmospheric muons such as the spallation products Lithium-9 and Helium-8 which perform a $\beta^-$-neutron decay [14, 15], fast neutrons which recoil in the liquid scintillator when entering the detector from the outside until they get captured [9], and stopping muons in coincidence with their Michel electrons/positron [16]. Additionally, random coincidences of backgrounds coming from different sources such as radioactive contamination exist [13, 17, 18].

## Details of the framework

The analysis approach is based on a binned maximum likelihood fit [19] of the energy-dependent event rate. We model our event rates in each bin by Poisson statistics. We describe the implementation of the framework in the following. For the source code see Ref. [4].

### General concept and statistical methods

For the likelihood approach, the expected number of neutrino events $n_i$ in each energy bin $1 \leq i \leq N$ is assumed to follow a probability density function (pdf), in case of continuous random variables, or a probability mass function (PMF), in case of discrete random variables, $f_i(n_i;\boldsymbol{\theta})$ that depend on parameters $\boldsymbol{\theta}$. In this analysis of discrete binned $n_i$ values the PMF is a Poissonian:

$$f_i = \frac{\mu_i^{n_i} e^{-\mu_i}}{n_i!}, \tag{3}$$

where the expectation $\mu_i(\vec{\theta})$ is a function of the model parameters. The joint probability $f$ over all $N$ energy bins $\vec{n} = (n_1, ..., n_N)^T$ is then given as:

$$f(\vec{n};\boldsymbol{\theta}) = \prod_{i=1}^{N} f_i(n_i;\boldsymbol{\theta}). \tag{4}$$

This pdf represents a likelihood function of the measured data vector $\vec{n}$ conditioned on the model parameters:

$$\mathcal{L}(\vec{n}|\,\vec{\theta}) = f(\vec{n};\vec{\theta}). \tag{5}$$

In the fit, this likelihood function is maximized with respect to the parameters $\boldsymbol{\theta}$ to find the most likely values $\hat{\boldsymbol{\theta}}$ given the measured data [20]. The targeted oscillation parameters $\vec{\eta} \in \vec{\theta}$ will be called signal parameters. The additional model parameters $\vec{\xi} \in \vec{\theta}$ that are fitted together with the signal parameters are called nuisance parameters. They typically include non-targeted oscillation parameters, normalizations of signal and background rates and parameters of the detector response and the reactor fluxes.

Usually several of the nuisance parameters and sometimes also signal parameters are constrained by a number $M$ of prior knowledge conditions $\vec{b}_j$ with $j \in M$ and $\vec{b}_j \in \vec{\theta}$. Those can constrain a single parameter or multiple parameters with correlations. This can be described by $M$ prior functions $g_j$ that express the pdfs of known constraints that are derived from independent measurements or known theoretical uncertainties. In case of uncorrelated conditions, the $g_j$ can be simple Gaussian functions $\mathcal{G}(b_j - \bar{b}_j, \sigma_{b_j})$ that describe the uncertainty of a model parameter $b_j$ where $\bar{b}_j$ and $\sigma_{b_j}$ are the expectation value and the standard deviation—or a more complicated pdf. Here, we use a generalized parametrization where the constrained parameters $\vec{b}_j(\vec{\theta})$ are functions of the actually fitted nuisance parameters $\vec{\theta}$. In case of correlations, the $g_j$ depend on the subset of $k$ correlated parameters $\vec{b}_j$ and the full covariance of the parameters $\vec{b}_j(\vec{\theta})$ is considered $\mathcal{G}((\vec{b}_j - \bar{\vec{b}}_j)^T \boldsymbol{\Sigma}_j^{-1} (\vec{b}_j - \bar{\vec{b}}_j))$, where $\boldsymbol{\Sigma}$ is the covariance matrix. The set of $M$ prior conditions is included in the likelihood function by multiplication:

$$\mathcal{L}(\vec{n}|\vec{\theta}) = f(\vec{n};\boldsymbol{\theta}) \cdot \prod_{j=1}^{M} g_j(\vec{b}_j(\vec{\theta})). \tag{6}$$

Often no prior function is applied to signal parameters $\vec{\eta}$ to avoid a biased measurement, however, if desired external prior information on the signal parameters can be consistently included in the fit.

Apart from finding the most likely value for the oscillation parameter $\theta_{13}$, or more general the signal parameters $\vec{\eta}$, it is also of interest to define a confidence interval for these parameter. We can achieve this in many situations by exploiting Wilks' theorem [21]: under certain regularity conditions and in the limit of infinite sample size, it holds that the statistic of two nested hypotheses $\Theta(\vec{\eta})$ and $\Theta_0(\hat{\vec{\eta}})$

$$\Lambda(\vec{n}, \vec{\eta}) := -2 \cdot \ln \frac{\sup\{\mathcal{L}(\vec{n}|\vec{\eta}, \tilde{\vec{\xi}})\}}{\sup\{\mathcal{L}(\vec{n}|\hat{\vec{\eta}}, \hat{\vec{\xi}})\}} \tag{7}$$

follows a $\chi^2$-distribution with the degrees of freedom given by the dimension of $\vec{\eta}$, i.e., $n_{dof} = 1$ in case of $\vec{\eta} = (\theta_{13})$. Here, $\tilde{\vec{\xi}}$ denotes the best-fit nuisance parameters for the tested values $\vec{\eta}$, i.e., the tested hypothesis $\Theta$, and $\hat{\vec{\xi}}$, $\hat{\vec{\eta}}$ are the global best-fit parameters, i.e., hypothesis $\Theta_0$. For finding an interval of confidence level $\alpha \in [0, 1]$ for $\vec{\eta}$ one has to find those values of $\vec{\eta}$ that correspond to the $1 - \alpha$ quantile of the $\chi^2$-distribution with the corresponding difference of degrees of freedom of the two hypotheses.

Similarly, the goodness-of-fit can be evaluated with a likelihood ratio:

$$\Lambda_S(\vec{n}, \vec{\eta}) := -2 \cdot \ln \frac{\sup\{\mathcal{L}(\vec{n}|\hat{\vec{\eta}}, \hat{\vec{\xi}})\}}{\mathcal{L}(\vec{n}|\vec{\mu} = \vec{n})}, \tag{8}$$

where the tested model becomes the global best fit, i.e., hypothesis $\Theta_0$ and the reference model is replaced by the saturated model $\Theta_S$ i.e., a model where the expectation is the experimental observation. The saturated model is thus describing the data perfectly and without any discrepancy [20]. Again, the likelihood ratio $\Lambda_S$ follows a $\chi^2$-distribution with the degrees of freedom given as dimensional difference between the saturated model $\Theta_S$ and the nested model $\Theta_0$. Note, that this difference is usually less than the difference in number of parameters, but can be determined by means of pseudo-experiments or ROOT's Minuit internal uncertainty estimation [22]. The goodness-of-fit is given by the p-value of the $\chi^2$-distribution as detailed above.

## Description of the input model of Double Chooz

An analysis may combine several datasets of different detector configurations. For the oscillation analysis in the context of Double Chooz that is not only data from the different phases of operation of the two detectors, but also phases where both nuclear reactors are turned on or off, or where only one or the other reactor is turned on.

The expected IBD events in each energy bin of all datasets are modeled by the sum of four contributions [9,

13]. Those contributions are the IBD-signal, as discussed in subsection Neutrino Signal, and the three background types (accidental, $\beta$-n decays of lithium and helium, and fast neutron backgrounds), as discussed in subsection Background processes. The three background spectra are determined from experimental data and thus enter the analysis without energy correction. Each of the three background contributions is then subject to modifications during the fit within their rate and shape uncertainties that are parametrized by nuisance parameters [23–25]. In contrast to those background spectra, the initial template for the IBD-signal spectrum is derived from a large statistics reactor and detector Monte-Carlo (MC) simulation that corresponds to the starting values of all fitted parameters. During fitting, changes of the fitted parameters are propagated to this expectation as detailed below. This includes uncertainties of the simulation, particularly the energy scale, flux uncertainties of the reactor simulation, normalization corrections, as well as the signal parameters $\sin^2(2\theta_{13})$ and $\Delta m^2_{ee}$. The background rate parameters and the signal normalization corrections scale the corresponding templates. Background shape parameters, and reactor flux parameters distort the templates in a bin-to-bin correlated way. In addition to the rate and shape parameters, three parameters for energy scale correction are fitted to match the simulated signal spectra with the data-driven background spectra [13].

The data of each dataset are binned for each detector between 1 MeV to 20 MeV with custom bin sizes resulting in a total of 38 energy bins. This allows for including shape information in the analysis. An exception is the dataset for time periods with both reactors turned off, where due to low statistics only the overall rate is fit [26]. Some energy bins are background-dominated. With this strategy to include background dominated regions in the fit, it is possible to constrain the background rates even beyond their model uncertainties [18, 23].

Most of the nuisance parameters describing the detector response, i.e., signal and background rates, energy scale and detection efficiencies are highly correlated between the datasets FD-I, ND, and FD-II and they are assumed fully correlated for phases of different reactor operation. A detailed discussion of correlations and their motivation can be found in reference [23].

The two signal parameters $\sin^2(2\theta_{13})$ and $\Delta m^2_{ee}$ are modeled with a full three flavor neutrino oscillation scenario including interactions of neutrinos with matter. The differential equations are solved with help of the GLoBES program [27, 28]. All other oscillation parameters are treated as nuisance parameters constrained to the world averages at the time.

## Likelihood implementation for Double Chooz within the framework

The model described in subsection Description of the input model of double chooz results in the following likelihood function [12, 23–25]:

$$
\begin{aligned}
\mathcal{L}(\vec{x}|\vec{\eta},\vec{\xi}) = &\prod_{i\in[E_{min}\cdots E_{max}]}\prod_{d\in\vec{d}} \mathcal{P}_i(n_{d,i},\mu_{d,i}(\vec{\eta},\vec{\xi})) \\
&\cdot \mathcal{P}_i(n_{\mathrm{off}},\mu_{\mathrm{off}}(\vec{\xi})) \\
&\cdot \prod_{j\in M}\mathcal{G}((\vec{b}_j-\vec{b}_{j,0})^T\mathbf{V}_j^{-1}(\vec{b}_j-\vec{b}_{j,0})).
\end{aligned} \tag{9}
$$

It is a product of multiplicative terms of the Poissonian likelihoods $\mathcal{P}_i(n_i,\mu_i)$ for the observed number of events $n_i$ in each energy bin $i$ for all considered datasets $\vec{d}$ from different phases of operation. For Double Chooz that is the dataset FD-I from the early phase of only far detector operation and the data ND of the near detector and FD-II of the far detector from the phase of simultaneous operation of both detectors. These phases are further split according to the combinations of reactors in operation. Here, $\mu_{d,i}(\vec{\eta},\vec{\xi})$ denotes the bin expectation, which is the sum of signal and background as a function of the model parameters. The second term is the Poisson probability of the total observed event rate during the reactor-off phases. The third term describes the $M$ Gaussian priors for constrained parameters. Note, that the notation includes the case of a single uncorrelated parameter as well as the possibility of different constraints for the different datasets. The oscillation parameters are the free signal parameters. While the parameter representing the expression $\sin^2(2\theta_{13})$ is left unconstrained, the value of the parameter representing the expression $\Delta m^2_{ee}$ is constrained with a prior based on independent measurements.

Systematic uncertainties are modeled by the following nuisance parameters $\vec{\xi}$ in the analysis:

- The normalizations of the reactor flux expectation for each energy bin are free fit parameters. With this approach the fit becomes independent of the initially modeled reactor flux prediction. The bin normalizations effectively become constrained by the data-to-data comparison between the measured data in each detector. This way, known discrepancies of reactor flux models cannot bias the fit, however, at the price of a slightly reduced sensitivity. The basis of the above approach is the high correlation of the observed reactor fluxes for the three datasets FD-I, FD-II, ND with almost-identical detectors. Because of different running times and small differences between the near and far detector, this assumption is only approximate. Therefore, we model additional constraints on the normalization of each energy bin of the three datasets with a total of $3 \times 41$ reactor flux parameters between 1 MeV to 11.25 MeV. The number of parameters is determined by the greatest common divisor of the bin widths to create a uniform binning. The bin content of the weighted and normalization corrected energy spectrum forms the basis of an area conserving spline, which is energy corrected and later transformed into the original binning. These parameters are correlated between the datasets with the above correlation factors and additionally we allow for uncorrelated shape deviations with a $41 \times 41$ covariance matrix for each dataset, that is determined from the reactor flux prediction.

- Backgrounds are modeled with free parameters for rate and shape. The shape of the contribution from rare isotopes ($^{9}$Li) is assumed identical between the three datasets. It is modeled with 38 shape parameters. The rate is assumed identical for FD-I and FD-II but is different for ND. Both total rates are not constrained but determined by the data as free parameters during the fit. The accidental backgrounds are modeled individually for each dataset by 38 parameters for the shape and one parameter for the rate. These parameters are assumed uncorrelated for the three datasets, to account for changes in data taking over time and differences in the detectors, but are constrained with a prior that reflects the uncertainty in the data-driven [18] determination of the rates. The spectrum of fast neutron and stopping muon backgrounds are modeled similar to the other backgrounds using a data-driven spectrum template. The shapes are assumed to be fully correlated between the datasets, while the rates are the same for FD-I and FD-II but independent for ND. A special case is the treatment of the small rate of $\overline{\nu}_e$ from the reactor fuel that has been determined during FD-I reactor-off phases [26]. As these neutrinos undergo the same oscillation as the signal, this is modeled consistently in the fit including the oscillated shape expectation and the rate is additionally constrained by a prior corresponding to this reactor-off rate.
- The uncertainties in the detector response are modeled by second-order polynomials [9]. They take into account the non-linearity of the visible energy response of the scintillator, the non-uniformity within the detector, and the charge non-linearity of the photomultiplier and electronics response. As we assume the energy response of the scintillator to be fully correlated but the other effects to be uncorrelated between the datasets, the 9 polynomial coefficients can be expressed by 7 independent parameters. In addition to the energy responses, the total detection efficiencies are subject to uncertainty. They are modeled by three constrained and partly correlated parameters. The uncertainty is dominated by the knowledge of the total target mass.

## Structure of the framework

The software framework `PhyLiNO` is designed to be modular and efficient, leveraging the advantages of modern features of the `C++` programming language up to `C++-20`. The minimization process uses the well-established `Minuit2` [22] module, part of the `ROOT` software framework [29]. Although the software framework relies for the minimization on `ROOT`, the internal sub-packages RooFit and RooStat are not utilized to increase the flexibility in the implementation. The `C++` framework `Eigen` [30] is used internally for linear algebra calculations, since it offers state-of-the-art performance coupled with flexibility in implementation and usage, but is not a fixed requirement for custom implementations. It is also possible to replace `Minuit2` by other minimizer algorithms with minimal changes to the interface. `PhyLiNO` implements all contributions to the likelihood function, which are subsequently used in the minimization process.

The basic structure of the fit procedure is illustrated in Fig. 1. In the center is the implemented likelihood function that calculates the Poisson and prior terms for the input data and model expectations. This function is minimized by iteratively modifying the model parameters until the estimated distance to the minimum is smaller than a predefined threshold which indicates that convergence is achieved.

The specific likelihood function and performance requirements differ between different experiments and are not easily generalizable. Therefore, the software framework is divided into three main components.

The first component manages the parsing of the parameters of a configuration file, that is designed to be easily extendable to each experiment's unique needs and thus

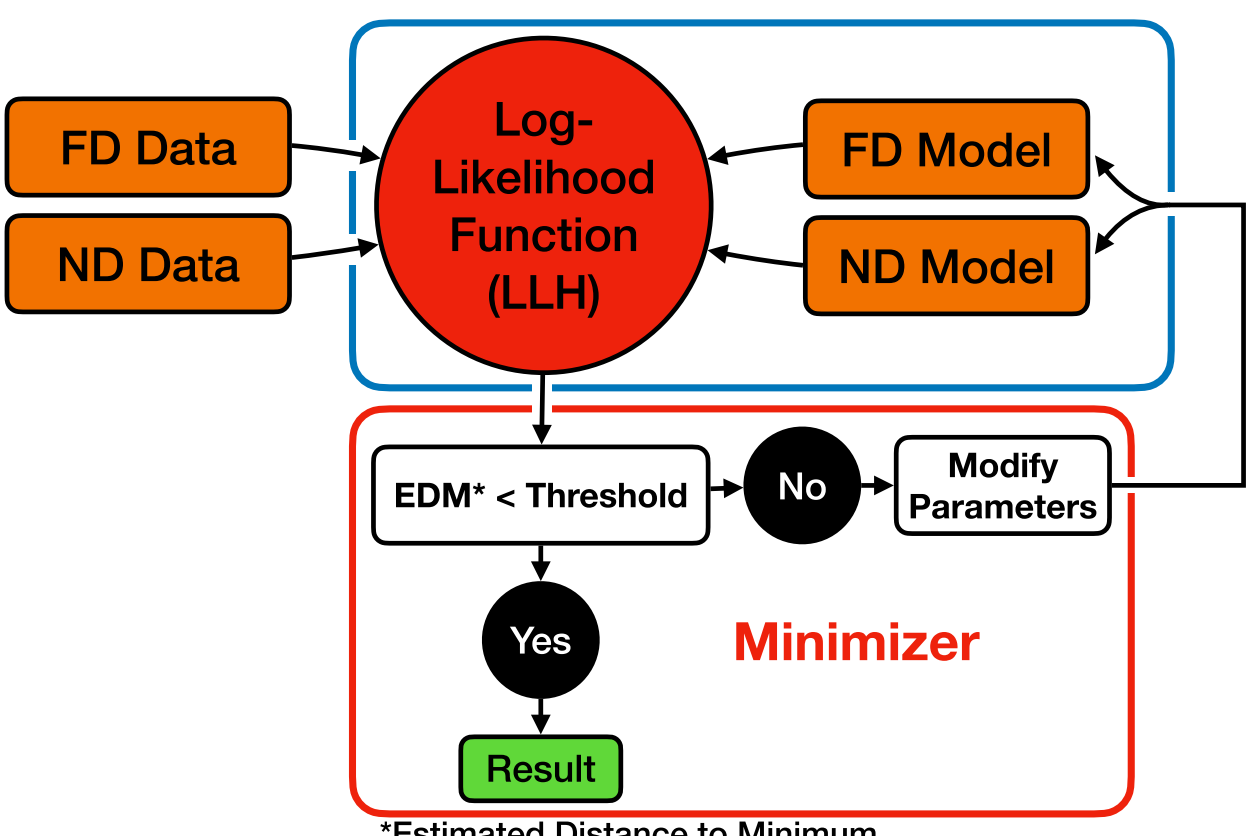

**Fig. 1** Basic structure of the fit procedure with the `PhyLiNO`-specific implementation highlighted with a blue box. The used `Minuit2` [22] minimizer is illustrated with a red box. `PhyLiNO` implements a structure to construct the parametrized model and calculate the likelihood for the data

allowing for flexible configuration across various use cases. This component also includes a custom C++ class that is utilized as a database to store data in memory and does not reload data for each computation step. This class significantly improves the performance of multiple parallel fits, e.g., for ensemble studies or parameter scans, by ensuring that frequently used information is readily available in memory without the need for repeated disk access. If the dataset becomes too large, alternatively usage of memory-mapped access could be used.

The second component implements the likelihood functions. It is highly modular. Likelihood functions are derived from abstract base classes with an interface for generalized parametrizations. Different contributions to the likelihood can be configured dynamically during execution time. The implementation includes features like Least-Recently Used (LRU) caching, where costly recalculations are only triggered upon relevant parameter changes larger than $1 \times 10^{-12}$. The necessary computation steps are modularized and thus different likelihood functions or even multiple experiments can be easily combined. The parameter handling allows for multiple namespaces and facilitates easy extension. The modular components are then integrated to form the overall model prediction, which is subsequently used to compute the likelihood.

The third component manages the output data. Once the fit is converged and the best-fit parameters are determined, they are saved in an output file formatted as JSON. This implements a standard interface to other programs of the analysis chain.

The Double Chooz implementation of the calculation of the likelihood is depicted in Fig. 2. All contributions to the full spectrum prediction, the background components, and the reactor neutrino signal are handled individually because they require specific treatments of the fit parameters and Monte Carlo datasets. The minimizer generically handles the minimized model parameters as uncorrelated. A transformation layer applies the covariance of any changed model parameter within the calculation of the likelihood.

The background expectations are based on data-driven templates that are dependent on the fit parameters. Each contribution is handled independently, with parameter LRU caching to optimize performance. The cached values are only recalculated when the parameters related to a specific contribution change, ensuring efficiency.

The signal expectation is derived from Monte Carlo events which include the full detector response. The calculation is divided into three distinct steps, each one is only recalculated if the respective parameters or the previous step have changed.

In the first step, these events are weighted to the modeled prompt energy spectrum, utilizing per-event weights depending on the oscillation formula given in Eq. 1. The events are sorted according to their energy and then each bin content is calculated sequentially to optimize memory access. Optionally, this step can be accelerated using GPUs, discussed in subsection GPU accelerated computation. In the next step, the bins of the energy spectrum, which already include the oscillation weighting, are rescaled to account for correlations. First, the correlations between the detector datasets and then the correlations between energy bins are applied. This approach captures correlated effects between the detectors while preserving only uncorrelated effects, such as neutrino oscillation, effectively absorbing deviations from the assumed reactor model. In the third step, an optional energy correction is applied, to account for small calibration discrepancies of the simulated and experimental data. The expectation is smoothed using a second-degree polynomial function. These corrected bins entries are then integrated over the original binning to generate the energy correction. This process effectively provides an adjustment to the energy reconstruction during execution time. As final step, all previously calculated signal and background contributions are combined for computing the likelihood.

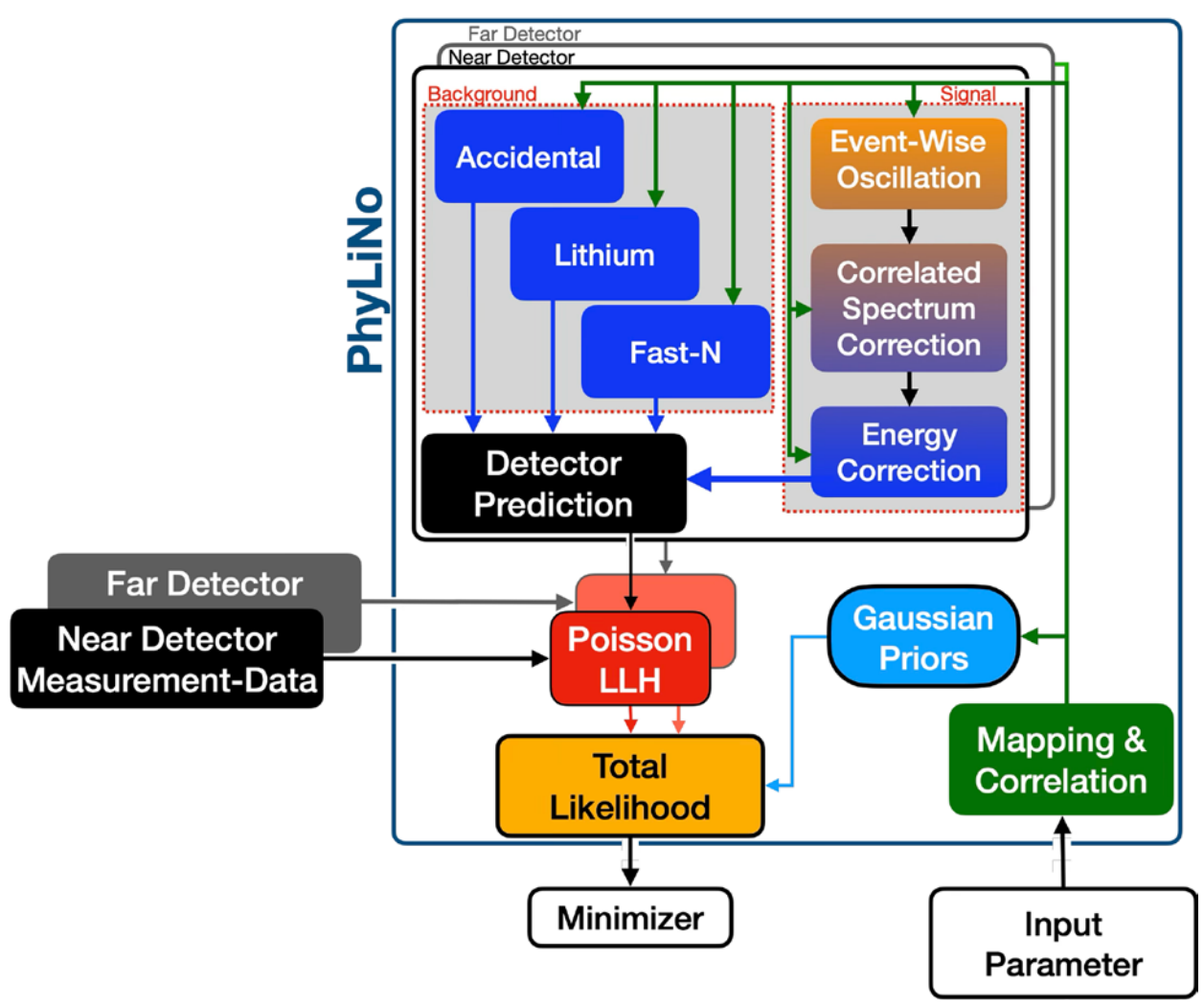

**Fig. 2** The specific likelihood implementation for the Double Chooz setup

Multiple fits can be parallelized with individual settings and initial parameter values. This allows for efficient parameter scans in profile-likelihoods. If user-implemented functions are thread-safe, parallelization can be achieved using threading. If not, multiprocessing with Message Passing Interface (MPI) can be used [31]. This parallelization approach is compatible with GPU calculations, provided the implementation is configured correctly.

### GPU accelerated computation

The calculation of the oscillation-weighted expectations for the large number of events in the simulation data is the most time-demanding computational step. However, the content of the different bins can be calculated in parallel, as they are independent of each other. The task is ideally suited for parallelization schemes such as multi-threading or GPU acceleration. In our implementation, only the computationally intensive neutrino oscillation function is replaced with a `CUDA`-accelerated version, which performs repeated evaluations with different inputs in parallel. This increases data throughput for repetitive computational tasks while leaving the remaining code base and the choice of minimizer unchanged. For example, the survival probability, defined in Eq. 1, depends only on two varying inputs energy and distance, while the neutrino oscillation parameters do not change within a single minimization step. The acceleration is implemented using `CUDA` [32] with `CUB` [33] from Nvidia. The implementation is integrated into a `CMake` configuration as a separate build option, allowing the framework to be built on machines without available GPU.

For the Double Chooz implementation, the reactor neutrino datasets for all detector configurations are copied to the GPU memory (VRAM) only once, as it remains unchanged throughout the minimization process. The resulting oscillated neutrino energy spectrum is copied back to the CPU memory for the following calculations. The calculation for each energy spectrum remains the same as previously described. The GPU acceleration reduces the total run-time as shown in Fig. 3. Note, that the performance gain depends on our used computing environment. The default fit takes roughly 39 s, which reduces to 15 s with multi-threading (factor $\sim 2.6$). Using a GPU further reduces the run-time to about 5 s (factor $\sim 3$). The total gain is a factor of $\sim 8$ in our setup [25].

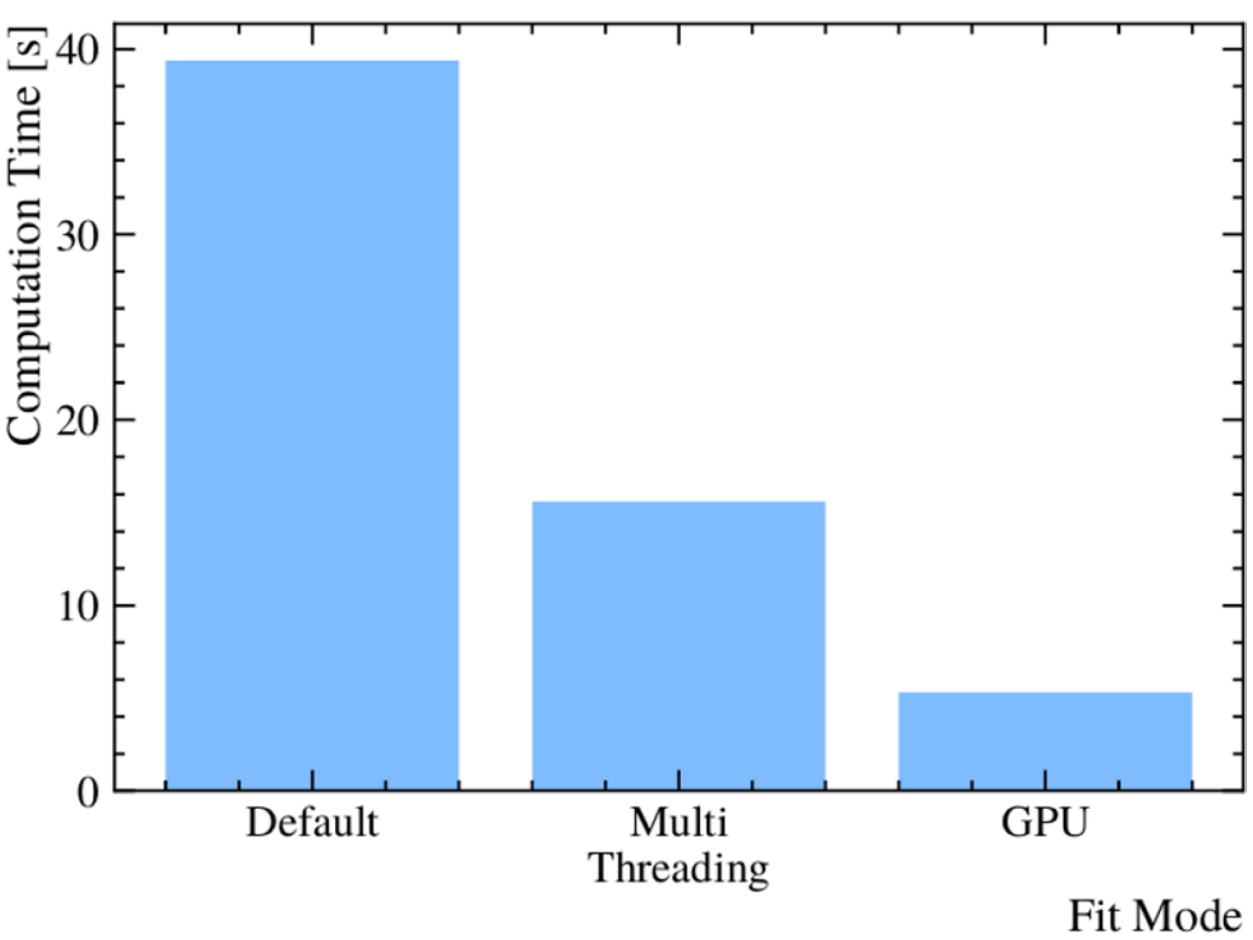

**Fig. 3** Comparison of total run-time for a fit of Double Chooz data for different computing environments. The computation platform is a single machine with a Nvidia RTX 3090 and an AMD Ryzen Threadripper 3960X with 48 threads. The default fit uses a single core; the multi-threading mode uses all 48 threads to calculate the individual event weights for the neutrino oscillation. The usage of a GPU for that task reduces the total fit time to about 5 s [25]

## Extensions of the framework

Due to its modular design, the framework has the ability to easily include additional datasets and their correlations. These can be data from the same experiment taken under different conditions or from a different experiment all together. This allows for a joint analysis of different datasets with a consistent treatment of systematic uncertainties and correlations. In the following, we discuss three examples of extended applications of `PhyLiNO`.

### Sterile neutrino searches

In Ref. [11, 12], the framework has been employed in the search for sterile neutrinos using data from Double Chooz. In the tested 3+1 model, two new fit parameters, the mass splitting and mixing angle with an additional fourth neutrino flavor have been introduced in the framework. The test statistic is unchanged from Eq. 7, but now the tested model parameters are $\vec{\eta} = \{\sin^2(2\theta_{14}), \Delta m^2_{41}\}$.

The nuisance parameters describing the reactor fluxes, detector responses, systematic uncertainties, and backgrounds are unchanged, but include the conventional oscillation parameters. A particular recent improvement [34] compared to the previous computational limitations in Ref. [11] is the high computing efficiency that now allows calculating more precise parameter scans with test statistics derived from large ensembles of pseudo-experiments. This high performance is generally relevant when evaluating test statistics for which Wilks' theorem is not applicable. Here, high performance allowed a full 2-D scan of the model parameters in a large mass range from $5 \times 10^{-4}\mathrm{eV}^2 \lesssim \Delta m^2_{41} \lesssim 5 \times 10^{-1}\mathrm{eV}^2$ with a large number of raster points.

### Conditional splitting of datasets

In many cases, the sensitivity of analyses can be improved by introducing additional information into the fit. Particularly, if background and signal expectations exhibit known variations splitting the total data into separate samples can be beneficial. This advantage has to be balanced with a potentially higher computational demand related to the higher dimensionality of the fit and additional nuisance parameters. As example, we discuss splitting the neutrino

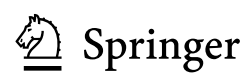

data in time according to different reactor configurations in Double Chooz [10, 25].

Neutrino oscillations depend on energy and traveled distance. If both reactor cores are active, the distance of detected neutrinos is ambiguous between the two, slightly differing baselines. However, if one of the two reactors is inactive, the baseline becomes unique. Splitting the data into three phases according to both and each single reactor operation results in a more predictive modeling and thus enhances the sensitivity for the measurement of $\sin^2(2\theta_{13})$. In the example of Double Chooz, this improvement is roughly 15 % [10, 25]. Multiple sets are treated consistently in our software framework by generically looping over all considered data.

As second example, the data can be separated also by the type of delayed neutron capture, i.e., capture on hydrogen or on gadolinium. The background predictions differ between the two capture-types and splitting the data includes additional information. Multiple splits strongly increase the complexity of the fit and the number of parameters, and therefore the high performance of the framework is key for pursuing such approaches. For these examples, the three analysis histograms (FD-I, FD-II and ND) become expanded to 18 histograms.

### Application to medium baseline reactor measurements

In order to verify the flexibility of the framework, we apply our framework to the test case of measuring the neutrino mass ordering with a medium-baseline neutrino detector setup similar to the JUNO experiment [35]. We assume a far detector located 53 km from the reactor source and a near detector at the Double Chooz ND baseline. For these baselines the survival probability of reactor-neutrinos becomes [36]:

$$P_{\bar{\nu}_e \to \bar{\nu}_e}(L,E) = 1 - \frac{1}{2}\sin^2(2\theta_{13}) \cdot \left(1 - \sqrt{\sin^2(2\theta_{12}) \cdot \sin^2(\Delta_{21})} \cdot \cos(\Delta_{ee} \pm \phi)\right) - \sin^2(2\theta_{12})\cos^4(\theta_{13})\sin^2(\Delta_{21}), \tag{10}$$

with $\Delta_{ab} \equiv \Delta m^2_{ab} \cdot \frac{L}{E}$, with $\Delta m^2_{ee}$ defined as in Eq. 2, and with $\phi \equiv \arctan\left(\cos^2(\theta_{12}) \cdot \tan(\Delta_{21})\right) - \Delta_{21}\cos^2(2\theta_{12})$. Depending on the mass ordering (normal or inverted) of the mass state 3 relative to 1 and 2, the sign of $\phi$ changes and thus modifies the fine-structure of the oscillating survival probability differently.

We generate an Asimov-type [3] pseudo-datasets of measured reactor-neutrino spectra in the two detectors based on the Double Chooz simulations scaled to the above baselines, detector sizes and smear the true energy with JUNO's energy resolution [35]. No systematic uncertainties and backgrounds are considered in this benchmark example. We fit the oscillation hypothesis to the data under the two different mass ordering assumptions. With the fit of the normal ordering assumption, we accurately retrieve the input parameters and thus validate the framework. The fit of the inverted ordering assumption demonstrates the framework's capability to distinguish between the two scenarios. This is presented in Fig. 4, which shows a profile of the obtained likelihood as a function of the parameter $\Delta m^2_{ee}$ for a lifetime of approximately 500 days. The likelihood of the inverted ordering is disfavored with respect to the injected normal ordering and the best fit $\Delta m^2_{ee}$ changes, as expected from the different value for inverted ordering.

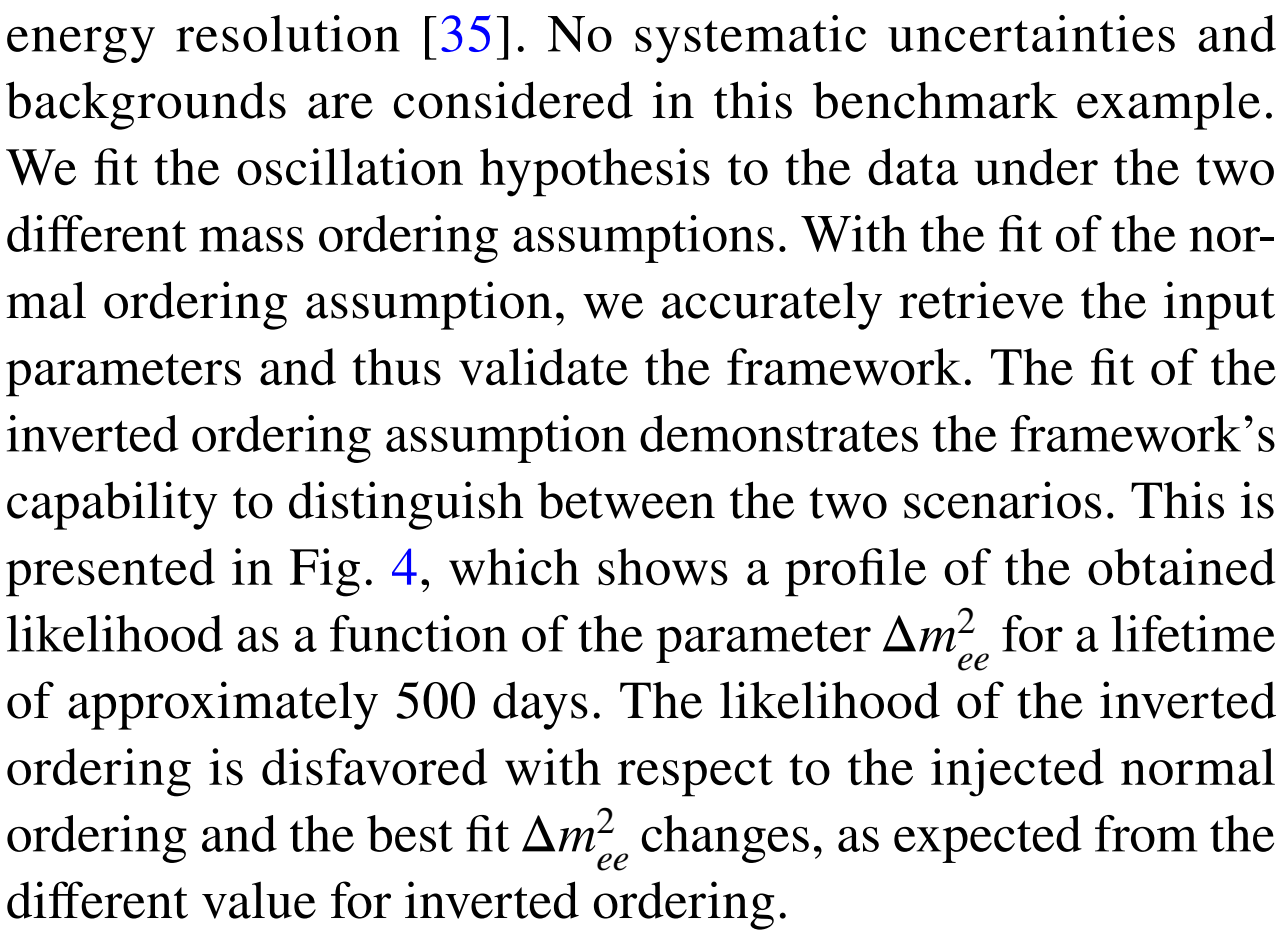

This example shows that the framework is easily adaptable to different experimental measurements. An important feature of the framework is also its ability to combine the likelihood functions from multiple experiments into a single global likelihood function. This approach would allow for a combined analysis of e.g., JUNO and other reactor neutrino experiments.

## Conclusion

In this article, we describe PhyLiNO, a forward-folding framework, which we developed within the context of the Double Chooz experiment. It complements existing tools developed at the same time in the context of hadron colliders [37, 38], through utilization of alternative implementations at several steps.

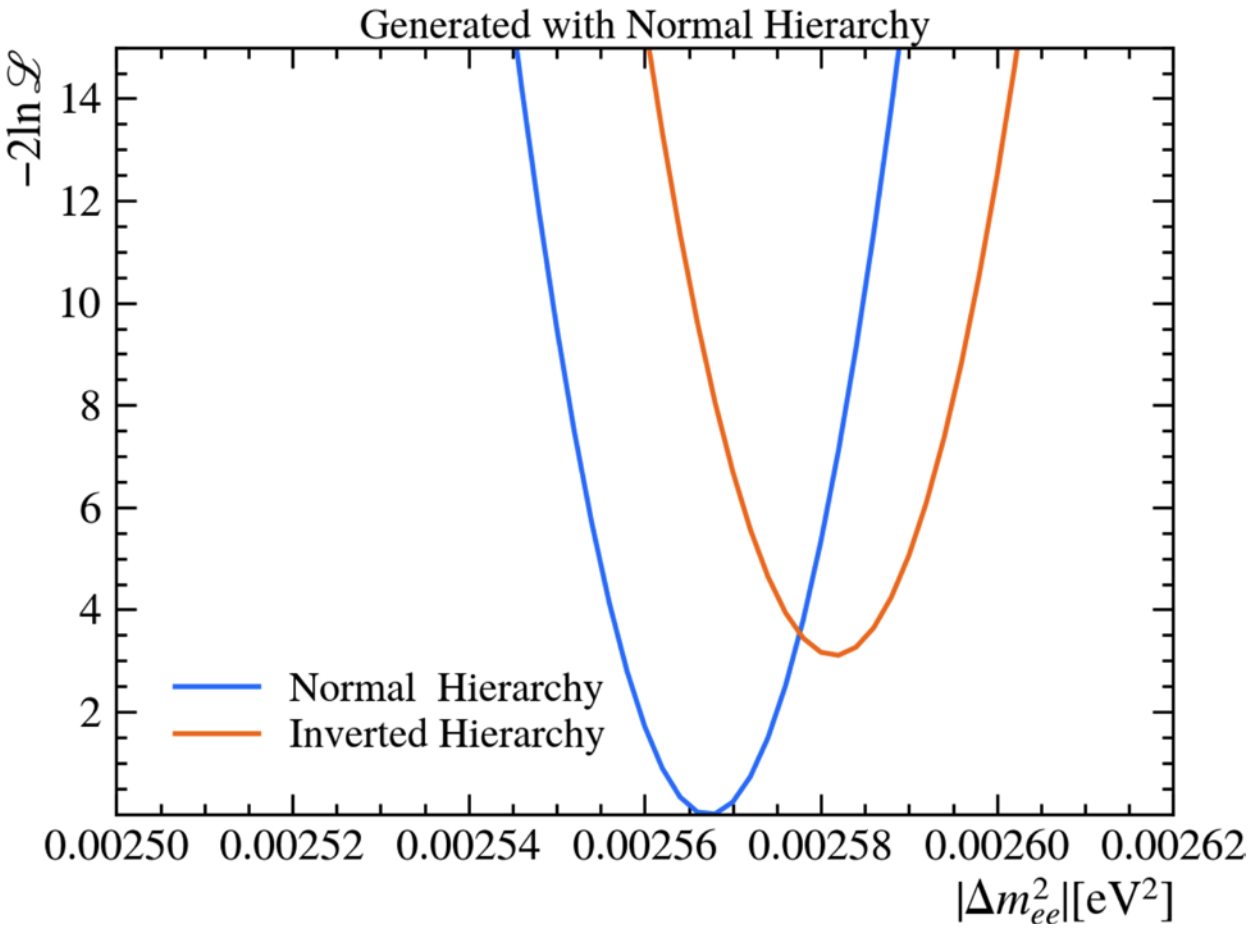

**Fig. 4** Profile likelihood scan for a generated pseudo-dataset of a medium baseline oscillation measurement with reactor-neutrinos. The blue curve (normal hierarchy) is the fit of the injected hypothesis of normal ordering, while the orange curve is the fit assuming inverted ordering. The shown likelihood is normalized to the global best likelihood

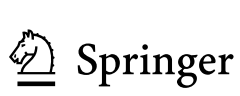

Our framework specializes on the analysis of neutrino oscillations and offers custom solutions for that task. It utilizes a binned likelihood approach. It enables a detailed treatment of systematics and their correlations. Due to its modular design, the framework is readily adaptable and expandable to handle data from different and even multiple experiments. Due to its included capabilities of parallelization, the framework can handle fit models with $\mathcal{O}(100)$ parameters and more with excellent computational performance. We have successfully deployed the framework in Double Chooz's analysis of three- and four-flavor neutrino oscillations. We further demonstrate its applicability to medium baseline reactor experiments.

**Acknowledgements** We sincerely thank the entire Double Chooz collaboration, especially its analysis group, led by Thiago Bezerra and Masaki Ishitsuka, for their critical input, thorough review, and validation of the fit framework. This work was supported by the German Research Foundation (DFG) through its Graduate College 1675: "Particle and Astroparticle Physics in the Light of LHC" (Project ID 164315326), the DFG Grant "Determination of the Leptonic Mixing Angle Theta-13 with the Double Chooz Experiment" (Project ID 39031316), as well as the Cluster of Excellence "Precision Physics, Fundamental Interactions, and Structure of Matter" (PRISMA$^+$ EXC 2118/1) funded by the DFG within the German Excellence Strategy (Project ID 390831469).

**Author contributions** S.S., P.S. and C.W. wrote the main manuscript text and P.S. prepared figures 1, 2, 3, 4. All authors reviewed the manuscript.

**Funding** Open Access funding enabled and organized by Projekt DEAL.

**Data availability** Data is provided within the supplementary files.

## Declarations

**Competing interests** The authors declare no competing interests.